\documentclass{aa}
\usepackage{graphicx}
\usepackage{txfonts}
\usepackage[colorlinks=true, linkcolor=blue, citecolor=blue, filecolor=blue, urlcolor=blue]{hyperref}
\begin{document}
\title{Can the giant planets of the Solar System form via pebble accretion in a smooth protoplanetary disc?}
\author{
Tommy Chi Ho Lau\inst{\ref{inst1},\ref{inst2}}
\and
Man Hoi Lee\inst{\ref{inst2},\ref{inst3}}
\and
Ramon Brasser\inst{\ref{inst4}}
\and
Soko Matsumura \inst{\ref{inst5}}
}

\institute{
University Observatory, Faculty of Physics, Ludwig-Maximilians-Universität München, Scheinerstr. 1, 81679 Munich, Germany
\label{inst1}
\and
Department of Earth Sciences, The University of Hong Kong, Pokfulam Road, Hong Kong
\label{inst2}
\and
Department of Physics, The University of Hong Kong, Pokfulam Road, Hong Kong
\label{inst3}
\and
Konkoly Observatory and Origins Research Institute, Research Centre for Astronomy and Earth Sciences; MTA Centre of Excellence; 15-17 Konkoly Thege Miklos ut, H-1121 Budapest, Hungary
\label{inst4}
\and
School of Science and Engineering, University of Dundee, Dundee, DD1 4HN, United Kingdom
\label{inst5}
}

\date{Received 2 September 2023; accepted 8 January 2024}

\abstract
{Prevailing $N$-body planet formation models typically start with lunar-mass embryos and show a general trend of rapid migration of massive planetary cores to the inner Solar System in the absence of a migration trap. This setup cannot capture the evolution from a planetesimal to embryo, which is crucial to the final architecture of the system.}
{We aim to model planet formation with planet migration starting with planetesimals of $\sim10^{-6}$ -- $10^{-4}M_\oplus$ and reproduce the giant planets of the Solar System.}
{We simulated a population of 1,000 -- 5,000 planetesimals in a smooth protoplanetary disc, which was evolved under the effects of their mutual gravity, pebble accretion, gas accretion, and planet migration, employing the parallelized $N$-body code SyMBAp.}
{We find that the dynamical interactions among growing planetesimals are vigorous and can halt pebble accretion for excited bodies. While a set of results without planet migration produces one to two gas giants and one to two ice giants beyond 6 au, massive planetary cores readily move to the inner Solar System once planet migration is in effect.}
{Dynamical heating is important in a planetesimal disc and the reduced pebble encounter time should be considered in similar models. Planet migration remains a challenge to form cold giant planets in a smooth protoplanetary disc, which suggests an alternative mechanism is required to stop them at wide orbits.}

\maketitle

\section{Introduction}\label{sec:intro}
Planet formation involves the growth from interstellar grains of sub-micron sizes to planets of thousands of kilometres in diameter, which is a process through at least 12 orders of magnitude in length scale. Details of the involved processes are still under ongoing research. Particularly, the formation of solid cores which subsequently accrete gas is a crucial yet still unclear step. This has been an active field of research for decades and requires further investigations.

\cite{Weidenschilling1977} presented a classic problem in planet formation that, due to aerodynamic drag in protoplanetary discs, solids of 10 cm to 1 m in size typically have a radial drift timescale of $\sim100$ years, which is much shorter than the typical disc lifetime of $1-10$ Myr. Furthermore, laboratory experiments of collisions \citep[e.g.][]{Wurm2005, GuettlerC.2010} also show a general behaviour that millimetre-sized grains require extremely small relative velocities to grow, so that fragmentation and bouncing are avoided. These barriers of particle growth are often summarized as the `metre-size barrier' in the literature. This implies that planetesimals of a kilometre in size have to form rapidly through the metre-sized scale from dust via an alternative process.

The Goldreich-Ward mechanism suggests the formation of planetesimals through gravitational collapse of a very dense dust disc as a result of dust settling \citep{Goldreich1973}, where the dust disc needs to be $\sim10^4$ times thinner than the gas disc. However, \cite{Cuzzi1993} showed that this cannot occur in a protoplanetary disc. The dense dust disc at the midplane, along with the gas in it, rotates at the Keplerian velocity; however, the gas disc immediately above rotates at a sub-Keplerian velocity due to the radial pressure gradient. This results in a steep vertical velocity gradient at the dust-gas interface, which induces the Kelvin-Helmholtz instability, preventing the dust disc from settling and collapsing gravitationally.

However, settling a dust disc with a solid density comparable to the gas density is possible without triggering the Kelvin-Helmholtz instability. Analyses in multiple works \citep[e.g.][]{Youdin2005,Youdin2007,Johansen2007,Johansen2009,Bai2010} suggest this can induce non-gravitational clumping of dusts due to disc turbulence or streaming instability. The over-densities of dust can subsequently collapse through gravity on an orbital timescale. Recent hydrodynamic numerical simulations \citep[e.g.][]{JohansenA.2012,Johansen2015,Simon2016,Simon2017} further show that dense filaments of solid particles undergo gravitational collapse and planetesimals up to about the size of Ceres are almost instantly formed. This process is a viable pathway for planetesimal formation.

The classical core accretion model of gas giant formation \citep{Mizuno1980,Pollack1996} requires a solid core of $\sim10 M_\oplus$. Beyond the critical mass, hydrostatic equilibrium in the gas envelope cannot be maintained, resulting in runaway gas accretion. The growth ends as the supply of gas is terminated due to gap opening in the disc or gas dispersal as the disc evolves.

Through $N$-body simulations, \cite{Kokubo1998,Kokubo2000} showed that pairwise accretion of planetesimals results in runaway growth, where more massive bodies grow faster. As protoplanets grow massive enough to interact with each other gravitationally, their orbital separations remain larger than $\sim5$ Hill radii and growth becomes oligarchic, where the growth rate is slower for more massive bodies. This results in a bimodal system of a few protoplanets and a population of small planetesimals. Their extrapolation estimates that the growth timescale to reach $5-10\ M_\oplus$ is of the order of $10-100$ Myr beyond $5$ au, which is much longer than the typical disc lifetime. Since a solid core of $\sim 10M_\oplus$ has to be formed before disc dispersal in order to accrete gas, a more efficient planetesimal growth mechanism is required.

Large populations of grains ranging from millimetres to tens of centimetres in radius, or pebbles, have been detected in protoplanetary discs by millimetre to centimetre observations \citep[e.g.][]{TestiL.2003, Wilner2005}. These observations are consistent with the metre-size barrier mentioned above. The growth of these small particles is stalled and they remain throughout most of the lifetime of the discs \citep{Cleeves2016}. This lays the foundation for the notion of pebble accretion. In this scenario, a large population of pebbles, as leftover solids, co-exists with planetesimals, in contrast to the classical scenario where pebbles are neglected for the growth of planetesimals of the order of a kilometre and beyond. Planetesimals that are massive enough to gravitationally deflect pebbles from the gas streamline and have a long enough encounter time can accrete a significant fraction of the drifting pebbles. This emerges as a mechanism for efficient planetesimal growth commonly called `pebble accretion' (\citealp{OrmelC.W.2010,LambrechtsM.2012,GuillotTristan2014}; \citealp[see][for review]{Johansen2017,Ormel2017}).

\cite{Kretke2014} conducted a series of numerical simulations incorporating pebble accretion with an initial mass spectrum of $\sim 10^6$ planetesimals. The Lagrangian Integrator for Planetary Accretion and Dynamics (LIPAD) \citep{Levison2012}, an $N$-body code, was deployed, which utilizes statistical algorithms to follow a large number of particles represented by tracers. As a result of oligarchic growth, the simulations generally form hundreds of $\sim M_\oplus$ bodies at $4-10$ au but further growth is halted due to gravitational scattering. The scattered oligarchs also pollute the inner Solar System with water and disrupt the outer Solar System.

To produce a Solar System analogue, the later work by \cite{Levison2015} modifies the pebble formation model that the pebble formation timescale is lengthened to $\sim 1$ Myr. This allows viscous stirring among planetesimals, which is on a shorter timescale compared to the growth timescale through pebble accretion. The less massive planetesimals are excited to orbit with higher inclinations. As the pebble density is lower farther away from the midplane of the disc, these inclined planetesimals are then starved of pebbles. This scenario yielded $1-4$ planets at $5-15$ au from the Sun without a stage of oligarchic growth. However, as noted in their work, gas accretion was cut off arbitrarily once the planet reaches the Jupiter mass $M_J$, instead of employing physical laws to stall the growth. Also, the embryos started to accrete gas in the simulations at around 8 Myr. The adopted gas accretion rate is likely unrealistically high as the disc has only $\sim4\%$ of its initial surface density at this age in their model, which results in a generous gas accretion rate. Finally, planet migration, which puts a critical time constraint on planet formation, was not considered in the model either.

\cite{MatsumuraSoko2017}, in turn, employed the Symplectic Massive Body Algorithm (SyMBA) \citep{Duncan1998}, a direct $N$-body code, with modifications to include pebble accretion, planet migration and gas accretion. They explored the dependence on stellar metallicity, stellar accretion rate and the viscosity parameter of the disc. Without migration, $1-3$ gas giants are formed at a few au in younger and less viscous discs. However, at the end of their 50 Myr simulations with migration, none of the results is consistent with the Solar System, as there are no giant planets left beyond 1 au. This shows that planet migration plays a crucial role in planet formation. Another major difference between the works by \cite{Levison2015} and \cite{MatsumuraSoko2017} is the number of particles simulated. \cite{Levison2015} use LIPAD, which simulates a large population of particles employing a statistical algorithm making viscous stirring among planetesimals possible. They also focused on growing gas giant analogous to the Solar System, and the domain of simulation is $4-15$ au. In contrast, \cite{MatsumuraSoko2017} focus on the production of the observed exoplanetary systems, and the domain of simulation is $0.3-5$ au instead.

More recently, \cite{BitschBertram2019} adopt the slower migration prescription in the high-mass regime by \cite{Kanagawa2018}. They employ the pebble and $N$-body code FLINTSTONE that also includes planet migration, eccentricity and inclination damping, as well as disc evolution. Their results show that with higher pebble mass flux and reduced planet migration rate, gas giants can indeed survive at wide orbits; with the final semimajor axes sensitive to the pebble mass flux and planet migration rate. Also, some of the resulting gas giants undergo scattering close to the Sun and end at a few au from the Sun. However, in these simulations, there are also other planets of a few to tens of $M_\oplus$ that migrate into the inner disc with less than 1 au, in contrast to the Solar System. Similarly, \cite{MatsumuraSoko2021} is able to form cold giant planets but cannot simultaneously avoid massive planetary cores migrating into the inner Solar System.

These works incorporating pebble accretion into global $N$-body simulations show intriguing results that the formation of gas accreting cores is possible through pebble accretion. Yet, further investigations are required to produce results that are consistent with the Solar System. The present study aims at assembling the giant planets analogous to those in the Solar System. In contrast to previous $N$-body planet formation models \citep[e.g.][]{MatsumuraSoko2017,BitschBertram2019,MatsumuraSoko2021} that focus on a small number of lunar-mass embryos, we assume an initial planetesimal disc with planetesimal sizes comparable to those formed via the gravitational collapse induced by streaming instability. This is made computationally possible by employing SyMBA parallelized (SyMBAp) \citep{Lau2023}, which is a parallelized version of SyMBA. In the following, Sect. \ref{sec:method} presents the methodology adopted in this work and the results are presented in Sect. \ref{sec:results}. The discussion of the results, the implications and caveats are in Sect. \ref{sec:dis}.

\section{Methods}\label{sec:method}
We generally follow the model by \cite{MatsumuraSoko2017} where additional subroutines are coupled with the symplectic direct $N$-body algorithm SyMBA \citep{Duncan1998} to study planet formation in a protoplanetary disc. To facilitate the integration of a self-gravitating planetesimal disc in this work, we instead employ SyMBAp \citep{Lau2023}. Further improvements are also made on the models of pebble accretion, gas accretion and the transition to the high-mass regime of planet migration. The following includes a summary of various parts of the model and the modifications made in this work are described in detail.

\subsection{Disc model}\label{sec:method_disc}
\begin{figure}
\centering
{\graphicspath{{./fig/}} \input{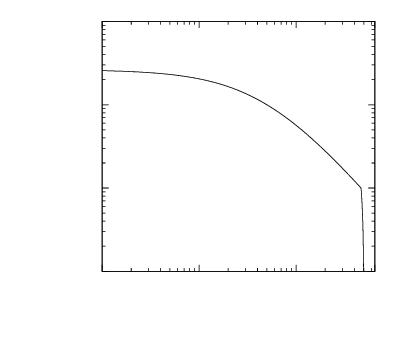}}
\caption{Time evolution of $\dot{M}_\ast$ with the initial age of the disc $t_0 = 0.5$ Myr. The value of $\dot{M}_\ast$ is turned down linearly when it drops below $10^{-9}M_\odot\,\text{yr}^{-1}$ to mimic the effect of photoevaporation.}
\label{fig:mdot_star}
\end{figure}
We consider an axisymmetric protoplanetary disc around a Solar-type star of $1 M_\odot$ in mass and $1 L_\odot$ in luminosity undergoing steady gas accretion. The gas accretion rate can be expressed as
\begin{equation}
	\dot{M}_\ast=3\pi\Sigma_\text{g}\nu \label{eq:gasacc}
\end{equation}
with $\Sigma_\text{g}$ the gas surface density. For the viscosity $\nu$, the \cite{Shakura1973} $\alpha$-parametrization is adopted such that
\begin{equation}
	\nu = \alpha_\text{acc} c_\text{s} H_\text{g} \label{eq:visc}
\end{equation}
with the viscosity parameter $\alpha_\text{acc}=10^{-3}$ set in this work. The isothermal sound speed is used and given by $c_\text{s} = \sqrt{k_\text{B}T/\mu}$ with the Boltzmann constant $k_\text{B}$, the disc midplane temperature $T$, the mean molecular weight of the gas $\mu=2.34m_\text{H}$, and the hydrogen mass $m_\text{H}=1.67\times10^{-27}$ kg. The gas disc scale height $H_\text{g}$ is defined by $H_\text{g}\equiv c_\text{s}/\Omega_\text{K}$, where the local Keplerian orbital frequency $\Omega_\text{K}=\sqrt{GM_\ast/r^{3}}$ with the gravitational constant $G$, the mass of the central star $M_\ast$, and the distance from the star $r$. Following \cite{Hartmann1998}, the evolution of the disc is propagated from the modulation of the stellar accretion rate by
\begin{equation}
	\log\left( \frac{\dot{M}_\ast}{10^{-8}M_\odot\,\text{ yr}^{-1}}\right) =-1.4 \log\left(\frac{t+t_0}{\text{Myr}}\right) \label{eq:stellacc}
\end{equation}
with the time since the start of the simulation $t$ and the initial age of the disc $t_0 = 0.5$ Myr. Fig. \ref{fig:mdot_star} shows the time evolution of $\dot{M}_\ast$. When $\dot{M}_\ast$ drops below $10^{-9}M_\odot\,\text{yr}^{-1}$, $\dot{M}_\ast$ is linearly turned down to zero at $t+t_0=5.5$ Myr to mimic the effect of photoevaporation following \cite{MatsumuraSoko2017}. With this setup, the initial stellar accretion rate is about $2.64\times10^{-8} M_\odot\,\text{ yr}^{-1}$ and reaches $10^{-9}M_\odot\,\text{yr}^{-1}$ when $t\approx4.68$ Myr. 

In general, the inner part of the disc is dominated by viscous heating and the outer part is dominated by radiative heating. Since this work focuses on the formation of the giant planets in the Solar System, only radiative heating is considered for the disc, in contrast to the disc model in \cite{MatsumuraSoko2017,MatsumuraSoko2021} where viscous heating is also considered. The midplane temperature profile of the disc $T$ is given by \citep{Oka2011}
\begin{equation}
    T= \ 150\ \left( \frac{r}{\text{au}}\right)^{-3/7} \text{ K}.\label{eq:T_rad}
\end{equation}
This setup yields the reduced disc scale height profile
\begin{equation}
    \hat{h}_\text{g}\equiv\frac{H_\text{g}}{r}\approx0.024\left(\frac{r}{\text{au}}\right)^{2/7}.\label{eq:hhat_rad}
\end{equation}
With Eq. (\ref{eq:gasacc}) for the gas accretion rate, Eq. (\ref{eq:visc}) for the $\alpha$-parametrization, and Eq. (\ref{eq:stellacc}) for the evolution of the stellar accretion rate, Eqs. (\ref{eq:T_rad}) and (\ref{eq:hhat_rad}) yield the gas surface density in the radiatively heated region
\begin{equation}
    \Sigma_\text{g}\approx2.7\times10^3 \left(\frac{\alpha_\text{acc}}{10^{-3}}\right)^{-1} \frac{\dot{M}_\ast}{10^{-8}M_\odot\text{ yr}^{-1}} \left(\frac{r}{\text{au}}\right)^{-15/14} \text{ g cm}^{-2}.
\end{equation}
This disc model yields a profile of the midplane pressure gradient parameter, where $P$ is the midplane gas pressure,
\begin{align}
    \eta &\equiv-\frac{\hat{h}_\text{g}^2}{2}\frac{\partial\ln P}{\partial\ln r}\nonumber\\
    &\approx8.02\times10^{-4}\left( \frac{r}{\text{au}}\right)^{4/7}.
\end{align}

\subsection{Planetesimal disc}
\begin{figure}
\centering
{\graphicspath{{./fig/}} \input{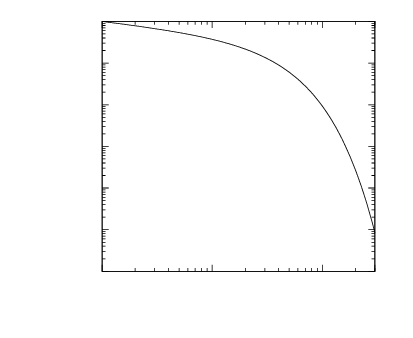}}
\caption{Adopted truncated power law initial planetesimal mass function as described by Eq. (\ref{eq:dN}) based on \cite{Abod2019}. It is presented in the unit of the planetesimal gravitational mass $m_G$.}
\label{fig:imf}
\end{figure}

\begin{figure}
\centering
\includegraphics{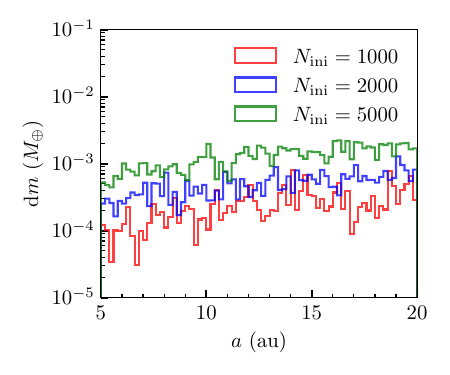}
\caption{Initial mass distribution of the realized planetesimal discs. One example is shown for each of the chosen values of $N_\text{ini}$. The width of each bin is 0.2 au.}
\label{fig:Sig_plts_ini}
\end{figure}

Instead of starting with lunar mass embryos as in \cite{MatsumuraSoko2017}, a planetesimal disc is generated from $5-20$ au initially with an initial mass function implemented in a manner similar to \cite{LauTommyChiHo2022} as summarized in the following. Planetesimals are drawn from the cumulative mass distribution in the work  on planetesimal formation by \cite{Abod2019}, which has the form of an exponentially truncated power law. The number fraction of planetesimals above mass $m$ is given by
\begin{equation}
	\frac{N_{>m}}{N_\text{ini}}=\left( \frac{m}{m_{\min}}\right) ^{-0.3}\exp\left( \frac{m_{\min}-m}{0.3m_G}\right), \label{eq:dN}
\end{equation}
for $m\geq m_{\min}$, with $m_{\min}$ being the minimum planetesimal mass considered, $N_{>m}$ is the number of particles with a mass $>m$, $N_\text{ini}$ is the initial number of particles, and $m_G$ is a planetesimal gravitational mass. We have set $m_{\min}=10^{-2}m_G$ in this work, which is well below the peak of the distribution of the planetesimal mass in each logarithm mass bin as noted by \cite{LauTommyChiHo2022}. The upper limit of $m$ is also artificially set at $3m_G$ in the realization algorithm to avoid a mathematical singularity. This value is an order of magnitude larger than the characteristic mass of the initial mass function ($0.3m_G$), where \cite{Abod2019} also show that the maximum planetesimal mass is about an order of magnitude more massive than the characteristic mass. In this manner, only an insignificant number of massive planetesimals ($\sim8\times 10^{-6} N_\text{ini}$) is lost. The form of the cumulative mass function is shown in Fig. \ref{fig:imf}. \\\

For $m_G$, we adopt the critical mass for gravitational collapse of a dust clump in the presence of turbulent diffusion by \cite{Klahr2020}, which is given by
\begin{align}
	m_G&=\frac{1}{9}\left( \frac{\delta}{\text{St}} \right)^{3/2} \hat{h}_\text{g}^{3} M_\odot  \label{eq:m_G}\\
	&\approx5.78\times10^{-4}\left( \frac{\delta}{10^{-5}}\right)^{3/2} \left(\frac{\text{St}}{10^{-2}} \right)^{-3/2} \left( \frac{\hat{h}_\text{g}}{0.038} \right)^3M_\oplus\nonumber
\end{align}
where $\delta$ is the small-scale diffusion parameter, which is independent of $\alpha_\text{acc}$, and $\text{St}$ is the Stokes number. In this work, we set $\delta=10^{-5}$ and $\text{St}=10^{-2}$ exclusively for planetesimal realization. While the strength of the small-scale diffusion is an active research topic in the field, the adopted value is motivated by the measurements of local diffusivity of dust particles in streaming instability presented in \cite{Schreiber2018}.

In each simulation, the semimajor axis $a$ of a new planetesimal is randomly drawn from $5-20$ au, which implies a surface number density of planetesimals that scales with $1/r$. The value of $m_G$ is then evaluated with the local disc scale height. Afterwards, the mass $m$ of this planetesimal is drawn from the mass function given by Eq. (\ref{eq:dN}) with the chosen value of $N_\text{ini}$ noted later in Sect. \ref{sec:num}. Figure \ref{fig:Sig_plts_ini} shows the initial mass distributions of the realized planetesimal discs with one example shown for each of the chosen values of $N_\text{ini}$. The eccentricity $e$ is randomly drawn from a Rayleigh distribution with the scale parameter $10^{-6}$. The inclination $i$ in radian is also drawn from a Rayleigh distribution but with the scale parameter $5\times10^{-7}$ instead. Other angles of the orbital elements are drawn randomly from $0$ to $2\pi$. The physical radius $R_\text{p}$ is calculated by assuming an internal density $\rho_\text{s}=1.5 \text{ g cm}^{-3}$. The realization process repeats until the total number of planetesimals reaches the chosen value. The planetesimals are then evolved under full gravitational interactions between themselves and the central star, as well as additional effects of pebble accretion (Sect. \ref{sec:intro_pa}), gas accretion (Sect. \ref{sec:ga}) and planet-disc interactions (Sect. \ref{sec:mig}).

\subsection{Pebble accretion}\label{sec:intro_pa}
\begin{figure}
	\centering
	{\graphicspath{{./fig/}} \input{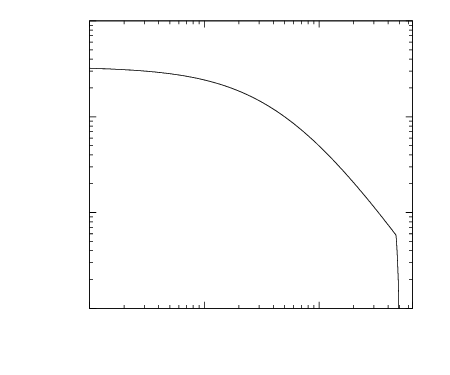}}
	\caption{Time evolution of the pebble mass flux $\dot{M}_\text{pf}$ given by Eq. (\ref{eq:pebflux}) with $Z_0=10^{-2}$ and  $\alpha_\text{acc}=10^{-3}$ as set in this work.}
	\label{fig:pebflux}
\end{figure}
We implement the `pebble formation front' model \citep{LambrechtsM.2014a} to estimate the pebble mass flux $\dot{m}_\text{peb}$. As dust particles coagulate and grow into pebbles, their velocities are strongly influenced by the headwind. This causes a significantly inward drift of pebbles that provide a solid mass flux to the inner part of the disc. Since the dust growth timescale increases with radius in general, the source of the pebble mass flux, or the pebble formation front, evolves outwards in time. The location of the pebble formation front $r_\text{pf}$ is given by \citep{LambrechtsM.2014a}
\begin{equation}
	r_\text{pf}(t)=\left( \frac{3}{16}\right) ^{1/3}(GM_\ast)^{1/3}(\epsilon_\text{d}Z_0)^{2/3}t^{2/3}
\end{equation}
with the initial dust-to-gas ratio $Z_0$ and the particle growth parameter $\epsilon_\text{d}=0.05$.
The pebble mass flux $\dot{M}_\text{pf}$ is then calculated from the dust mass swept across by the pebble formation front per unit time, that is,
\begin{align}
	\dot{M}_\text{pf}&=2\pi r_\text{pf}Z_0\Sigma_\text{g}(r_\text{pf})\dot{r}_\text{pf}\nonumber\\
	&\approx\frac{\dot{M}_\ast}{10^{-8}M_\odot\text{ yr}^{-1}}
    \left(\frac{Z_0}{10^{-2}}\right)^{5/3}
    \left(\frac{\alpha_\text{acc}}{10^{-3}}\right)^{-1}\times\nonumber\\
    & \ \ \ \ \ \ \left(\frac{t+t_0}{\text{Myr}}\right)^{-1/3}10^{2}M_\oplus\text{ Myr}^{-1}.\label{eq:pebflux}
\end{align}
A factor of $r^{-1/14}_\text{pf}$ is omitted for simplicity. We set $Z_0=10^{-2}$ in this work and Fig. \ref{fig:pebflux} shows the time evolution of $\dot{M}_\text{pf}$ for the chosen parameters. We note that at 4.5 Myr, briefly before disc dispersal, $r_\text{pf}\approx350\text{ au}$.
This is comparable to the typical observed disc sizes, which is of the order of 100 au \citep[e.g.][]{Andrews2018,Long2018,Cieza2021}. In \cite{MatsumuraSoko2017}, $\dot{M}_\text{pf}$ is halved inside of the snow line. However, this treatment is not implemented in the present work as it focuses on the outer Solar System where particles are removed before they can reach the ice line in our model. The radial domain of this work is summarized later in Sect. \ref{sec:num}. On the other hand, we follow \cite{MatsumuraSoko2021} and adopt the pebble disc scale height given by
\begin{equation}
    H_\text{peb}=\left(1+\frac{\text{St}}{\alpha_\text{turb}}\right)^{-1/2}H_\text{g}
\end{equation}
with the Stokes number of pebble St. Following \cite{Ida2018}, an $\alpha_\text{turb}$ parameter is introduced, which is about an order of magnitude smaller than $\alpha_\text{acc}$ as evaluated by \cite{Hasegawa2017}. The latter is distinct from that in the classical $\alpha$-parametrization, i.e. the $\alpha_\text{acc}$ parameter introduced in Sect. \ref{sec:method_disc}. In this work, we set $\alpha_\text{turb}/\alpha_\text{acc}=0.1$. The $\alpha_\text{turb}$ parameter is also used for prescribing gas accretion (Sect. \ref{sec:ga}) and planet-disc interactions (Sect. \ref{sec:mig}) as described in the respective sub-sections.

Furthermore, the pebble flux available to each body is subtracted by the total pebble accretion rate of the superior bodies that are farther from the central star, if there are any. We define a pebble accretion efficiency $\epsilon_\text{PA}$ such that the growth rate of a body $i$ by pebble accretion is given by
\begin{equation}
	\dot{m}_{\text{PA},i}=\epsilon_{\text{PA}} \max\left( \dot{M}_\text{pf}-\sum^{N}_{n=i+1}\dot{m}_{\text{PA},n},0 \right),
\end{equation}
where bodies $(i+1)$ to $N$ are all the superior ones.

In this work, we also compare the pebble accretion efficiency of \cite{IdaS.2016} with modifications by \cite{MatsumuraSoko2021}, $\epsilon_\text{IGM16}$, and that by \cite{LiuBeibei2018} and \cite{OrmelChrisW2018}, $\epsilon_\text{OL18}$. In the derivation of $\epsilon_\text{IGM16}$, the pebble-accreting body is assumed to be in a circular orbit as noted in Sect. 3.2 of \cite{IdaS.2016}  and shown in Eq. (33) of their work regarding the pebble relative velocity. In contrast, \cite{LiuBeibei2018} and \cite{OrmelChrisW2018} do not hold this assumption, and both the inclination and the eccentricity of the pebble-accreting body contribute to the pebble relative velocity. The modifications of $\epsilon_\text{IGM16}$ made by \cite{MatsumuraSoko2021} considered the inclination of the body. However, it only plays a role in the calculation of the pebble volume density as shown in Eq. (32) of their work but not in the calculation of the pebble relative velocity. The differences between the two pebble accretion prescriptions and the consequences are further discussed in Sect. \ref{sec:dis_pa}.

When the planetesimals grow into massive cores, the process of pebble isolation occurs when they perturb the gas surface density profile and stop pebbles from reaching the planet itself as well as the inferior bodies that are closer to the central star, if there are any. We follow the assumption in \cite{MatsumuraSoko2017} that the required mass, which is often called the `pebble isolation mass', is given by
\begin{align}
	m_\text{iso}&=\frac{1}{2}\hat{h}_\text{g} ^3M_\ast \\
	& \approx 9.14 \left( \frac{\hat{h}_\text{g}}{0.038} \right)^3M_\oplus \nonumber.
\end{align}
Once any planet reaches this mass, pebble accretion is stopped for this planet and all the inferior ones if there are any.

\subsection{Gas accretion}\label{sec:ga}
When a massive core has formed and its solid accretion rate is low, gas can contract and form an envelope. We follow \cite{Ikoma2000} for the critical mass for runaway gas accretion, which is given by, for planet $i$,
\begin{equation}
	m_\text{g,crit}=10 \left( \frac{\dot{m}_{\text{PA},i}}{10^{-6} M_\oplus \text{ yr}^{-1}} \frac{\kappa}{1 \text{ cm}^{2} \text{ g}^{-1}} \right) ^p M_\oplus.
\end{equation}
In this work, we set the parameter $p=0.25$ \citep{Ida2004} and the envelope opacity $\kappa=1\text{ cm}^{2} \text{ g}^{-1}$. For cores that have reached this mass, we assume the gas envelope collapses on the Kelvin-Helmholtz timescale $\tau_\text{KH}$ given by \citep{Ikoma2000, Ida2004}
\begin{equation}
	\tau_\text{KH}= 10^9 \left( \frac{m}{M_\oplus} \right) ^{-3} \left( \frac{\kappa}{1 \text{ cm}^{2} \text{ g}^{-1} }\right) \text{ yr}.
\end{equation}
There are two factors that limit the actual gas accretion rate considered in our model. First, the gas supply is limited by the stellar accretion rate as well as the gas accreted by the superior planets. Also, gap opening by the planet shall further limit the gas accretion rate. And, we assume gas accretion is exponentially cutoff when the planet's Hill radius equals the local disc scale height, which is given by $m_\text{Hill}=3M_\ast\hat{h}^3_\text{g}$. These can be summarized as the expression for the gas accretion rate of planet $i$
\begin{align}
	\dot{m}_{\text{g},i}=\min \Biggl[ &\frac{m}{\tau_\text{KH}}, \nonumber\\
        &\max \left( \dot{M}_\ast-\sum_{n=i+1}^{N}\dot{m}_{\text{g},n} \ ,0\right) f_\text{local}\exp\left( -\frac{m}{m_\text{Hill}}\right) \Biggr]
\end{align}
where planets $(i+1)$ to $N$ are all the superior ones and the reduction factor $f_\text{local}$ is given by \citep{Ida2018}
\begin{equation}
	f_\text{local}=\frac{0.0308\hat{h}^{-4}_\text{g}(m/M_\ast)^{4/3}\alpha_\text{acc}^{-1}}{1+0.04K}.
\end{equation}
The gap opening factor $K$ is given by Eq. (\ref{eq:K}) in the next subsection (Sect. \ref{sec:mig}).

\subsection{Planet-disc interactions}\label{sec:mig}
Other than the $N$-body gravitational interactions, the bodies also experience the torques due to the planet-disc interactions. We adopt the prescription based on dynamical friction by \cite{Ida2020} and the transition from the low-mass to the high-mass regime by \cite{Ida2018} based on the gap opening factor $K$ by \cite{Kanagawa2015}. The timescales for the non-isothermal case and finite inclination $i$, while $i<\hat{h}_\text{g}$, (Appendix C and D of \citealp{Ida2020} and \citealp{MatsumuraSoko2021}) are implemented. The evolution timescales of semimajor axis, eccentricity and inclination are defined respectively by
\begin{equation}
	\tau_a\equiv-\frac{a}{\text{d}a/\text{d}t},\tau_e\equiv-\frac{e}{\text{d}e/\text{d}t}, \tau_i\equiv-\frac{i}{\text{d}i/\text{d}t}.
\end{equation}
These timescales are given by
\begin{align}
	\tau_a = \frac{t_\text{wav}'}{2\hat{h}_\text{g}^2} \Biggl[ &\frac{\Gamma_L}{\Gamma_0}\left(1-\frac{1}{C_\text{M}} \frac{\Gamma_L}{\Gamma_0} \sqrt{\hat{e}^2+\hat{i}^2}\right) ^{-1} + \nonumber\\
                &\frac{\Gamma_C}{\Gamma_0}\exp\left( -\frac{\sqrt{\hat{e}^2+\hat{i}^2}}{e_\text{f}}\right) \Biggr]^{-1},
\end{align}
\begin{equation}
	\tau_e =1.282t_\text{wav}'\left[ 1+ \frac{(\hat{e}^2+\hat{i}^2)^{3/2}}{15} \right],
\end{equation}
\begin{equation}
	\tau_i =1.838t_\text{wav}'\left[ 1+ \frac{(\hat{e}^2+\hat{i}^2)^{3/2}}{21.5} \right] ,
\end{equation}
where $\hat{e}\equiv e/\hat{h}_\text{g}$, $\hat{i}\equiv i/\hat{h}_\text{g}$, and we follow \cite{Fendyke2014} for the factor $e_\text{f}=0.01+\hat{h}_\text{g}/2$. The normalized Lindblad torque $\Gamma_L/\Gamma_0$ and corotation torque $\Gamma_C/\Gamma_0$ are described in detail by \cite{Paardekooper2011}. The characteristic time including the transition to the high-mass regime $t_\text{wav}'$ \citep{Tanaka2002,Ida2018} is given by
\begin{equation}
	t_\text{wav}'=\left( \frac{M_\ast}{m}\right) \left( \frac{M_\ast}{\Sigma_{\text{g}}r^2}\right) \left( \frac{\hat{h}_\text{g}^4}{\Omega_\text{K}}\right)\left(1+0.04K\right)
\end{equation}
with the gap opening factor $K$ given by
\begin{equation}
	K=\left( \frac{m}{M_\ast}\right)^2\hat{h}^{-5}_\text{g}\alpha_\text{turb}^{-1}. \label{eq:K}
\end{equation}
As noted in \cite{LauTommyChiHo2022}, it is more suitable to evaluate the value of $\Omega_\text{K}$ at the instantaneous distance from the star $r$ of the body instead of its semimajor axis $a$ in $N$-body simulations with large number of particles due to potential frequent encounters. We follow \cite{Ida2018} and introduce the $\alpha_\text{turb}$ parameter set to $\alpha_\text{turb}/\alpha_\text{acc}=0.1$ as described in Sect. \ref{sec:intro_pa}. The three timescales are applied to the equation of motion
\begin{equation}
	\mbox{\boldmath $a$}=-\frac{v_\text{K}\cdot S_a}{2\tau_a}\mbox{\boldmath $e$}_\theta-\frac{v_r}{\tau_e}\mbox{\boldmath $e$}_r-\frac{v_\theta-v_\text{K}}{\tau_e}\mbox{\boldmath $e$}_\theta-\frac{v_z}{\tau_i}\mbox{\boldmath $e$}_z
\end{equation}
in the cylindrical coordinates $(r,\theta,z)$ with the velocity of the embryo $\mbox{\boldmath $v$}=(v_r,v_\theta,v_z)$ and the local Keplerian velocity $v_\text{K}=r\Omega_\text{K}$. A switch for planet migration $S_a$ is introduced to toggle the evolution of the semimajor axis, which is turned off and on respectively by setting $S_a$ to $0$ and $1$ in this work.

\subsection{Numerical setups}\label{sec:num}
To explore the dependence on the total number of planetesimals, three values of $N_\text{ini}=\{1000,2000,5000\}$ are chosen. They translate respectively to a total planetesimal mass of about $\{0.02,0.04,0.1\}M_\oplus$. We test two pebble accretion efficiency prescriptions $\epsilon_{\text{PA}}=\{\epsilon_\text{IGM16},\epsilon_\text{OL18}\}$ described in Sect. \ref{sec:intro_pa} and the two states of $S_a=\{0,1\}$ described in Sect. \ref{sec:mig} that switches off or on the evolution of semimajor axis due to planet-disc interactions. Each simulation lasts for 6.5 Myr to allow for further dynamical evolution due to gravitational interactions after disc dispersal. Particles are removed if the heliocentric distance is less than 1 au or greater than 100 au. For each combination of the parameters, we conduct five simulations to sample the stochastic variations in the outcome. Thus a total of 60 simulations are conducted in this work and presented in the next section.

\section{Results}\label{sec:results}
The first part of this section (Sect. \ref{sec:migoff}) presents the results with migration turned off, i.e. $S_a=0$, followed by Sect. \ref{sec:migon} where the results with migration turned on, i.e. $S_a=1$, is presented.

\subsection{Simulations without planet migration ($S_a=0$)}\label{sec:migoff}
\subsubsection{Pebble accretion efficiency $\epsilon_{\text{PA}}=\epsilon_\text{IGM16}$} \label{sec:igm16}
\begin{figure*}
	\centering
	\includegraphics[width=\textwidth]{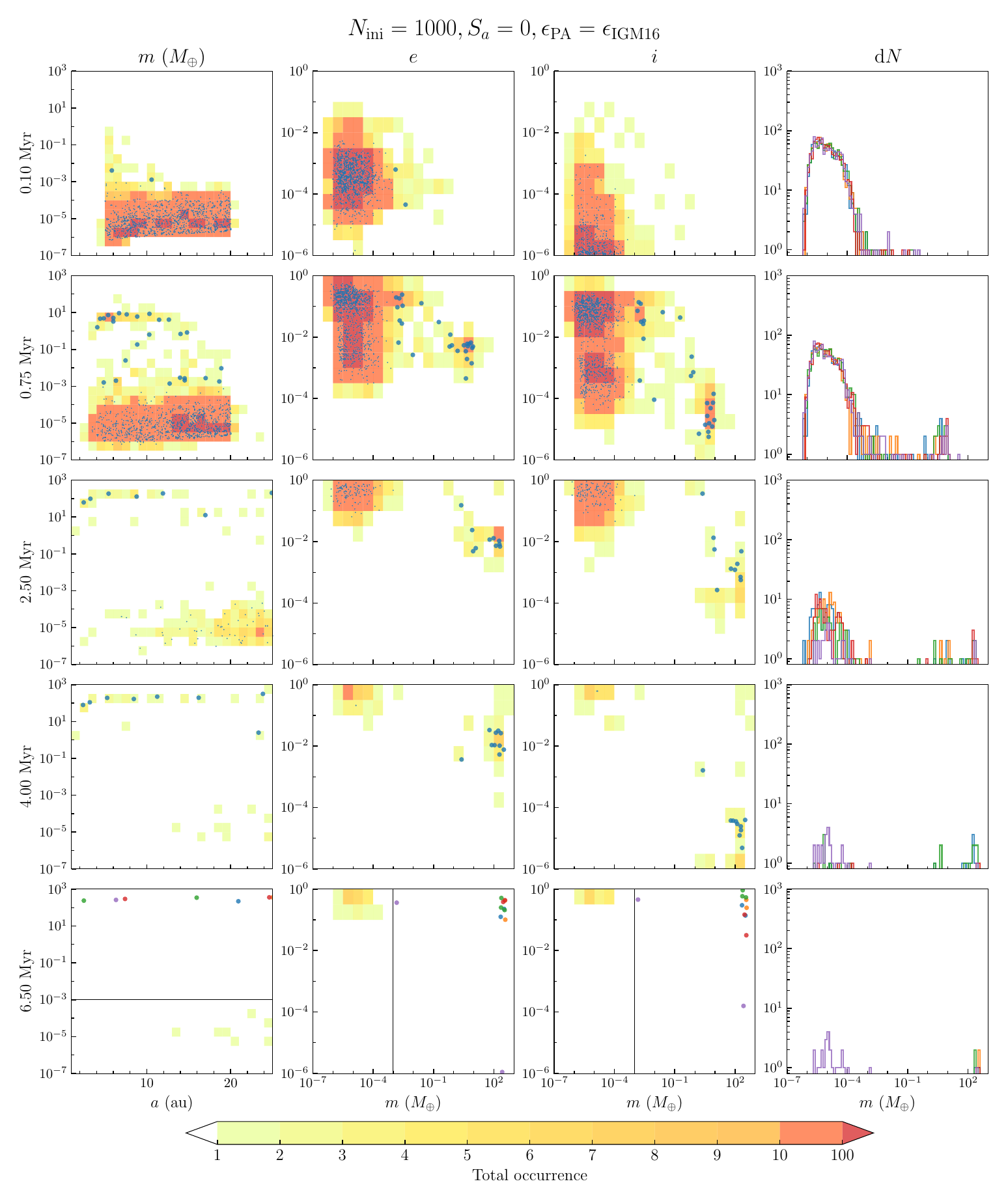}
	\caption{Results for the simulations with $N_\text{ini}=1000$, migration turned off ($S_a=0$), and the pebble accretion efficiency $\epsilon_{\text{PA}}=\epsilon_\text{IGM16}$. Each row presents a snapshot of the simulations at the time indicated by the timestamp on the left. For the first three columns from the left, the total occurrences of particles across all five random simulations are shown by heat maps with $2\times2$ cells in each minor axis grid cell. The left-most column shows $m$ and $a$. The next two columns to the right respectively shows the $e$ and $i$, respectively, against $a$. The right-most column shows the differential mass distribution of all bodies with each colour corresponds to one of the five simulations. Particles in one of the five simulations (blue) is also plotted with enlarged dots denoting particles above $10^{-3}M_\oplus$. For 6.5 Myr, i.e. the end of the simulations, particles above $10^{-3}M_\oplus$ in all five simulations are shown individually without using heat maps. Further descriptions are in the text.
	}
	\label{fig:1k_mig_off_IGM16}
\end{figure*}
\begin{figure}
	\centering
	\includegraphics{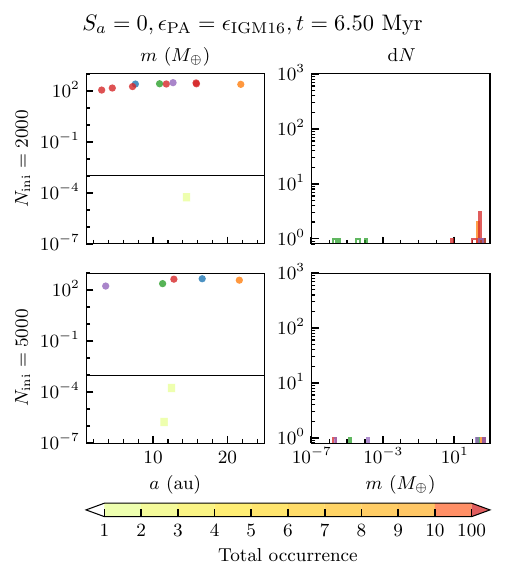}
	\caption{End results for the simulations with $N_\text{ini}=2000$ and $5000$, respectively, as indicated on the left, migration turned off ($S_a=0$), and the pebble accretion efficiency $\epsilon_{\text{PA}}=\epsilon_\text{IGM16}$. The two columns here correspond to the left-most and the right most columns of Fig. \ref{fig:1k_mig_off_IGM16}, respectively. There is no qualitative difference in the end results among the simulations with the chosen set of $N_\text{ini}=\{1000,2000,5000\}$.}
	\label{fig:2k_5k_mig_off_IGM16}
\end{figure}

Figure \ref{fig:1k_mig_off_IGM16} shows the results for $N_\text{ini}=1000$, $S_a=0$ and $\epsilon_{\text{PA}}=\epsilon_\text{IGM16}$. Each row presents a snapshot of the simulations at $t=\{0.10,0.75,2.50,4.00,6.50\}$ Myr respectively. For the first three columns from the left, the total occurrences of particles across all five simulations are shown by heat maps. The left-most column shows the mass $m$ in $M_\oplus$ and the semimajor axis $a$. The next two columns to the right show the eccentricity $e$ and inclination $i$ against $m$ respectively. The right-most column shows the differential mass distribution of the particles with each colour corresponds to one of the five simulations. Particles in one of the five simulations (blue) is also plotted with particles above $10^{-3}M_\oplus$ denoted by enlarged dots. For the last row (6.5 Myr), which shows the end results, particles above $10^{-3}M_\oplus$ in all simulations are shown individually (with a different colour for each simulation) without using heat maps.

The $m$--$a$ plots show a rapid growth by pebble accretion in the inner part of the disc in the first 0.1 Myr of the simulations. Some planetesimals in the massive tail of the distribution have grown by more than 3 orders of magnitude dominantly by pebble accretion. The growth rate has a strong dependence on the distance from the star, and particles closer to the central star accrete pebble much faster, as predicted by \cite{IdaS.2016}. This is also consistent with the analysis which includes both pebble and planetesimal accretion in \cite{Coleman2021}, though our simulations focus on the outer Solar System.

The $e$--$m$ plots and the $i$--$m$ plots show the early and fast growing bodies quickly heat up their neighbouring planetesimals from the beginning of the simulations to 0.75 Myr, increasing the eccentricities and inclinations of neighbouring planetesimals. The massive cores of $\sim M_\oplus$ stop further growth of the neighbouring smaller bodies by viscous stirring, with about 20 bodies having reached $\sim1-10M_\oplus$ by $0.75$ Myr. This effect of viscous stirring on pebble accretion is consistent with \cite{Levison2015} and further discussed in Sect. \ref{sec:dis_pa}. The $e$ and $i$ of these cores are also damped and remain low in contrast to those of the smaller bodies, which allows these massive bodies to further increase in mass due to the proximity to the dense pebble disc. This effect is more noticeable from the differential mass distributions, i.e. the rightmost column, that only the particles in the massive tail of the initial planetesimal population can grow significantly while the rest remain about the same mass. The growth of these massive bodies is drastically different from the traditional oligarchic growth scenario, where the growth is slowed down by viscous heating that clears nearby planetesimals. Here, the more massive bodies can continue growth via pebble accretion until reaching the pebble isolation mass, which is a result of the perturbations to the gas disc.

As the simulations progress forward, the massive cores grow further by gas accretion and eject most of the small bodies from $0.75-4$ Myr. At the end of the simulations, i.e. $t=6.50$ Myr, some of the massive cores and gas giants ($m>10^2M_\oplus$) formed have been ejected, and $1-4$ gas giants remain but their locations vary greatly across the simulations. This indicates a strong stochastic behaviour due to dynamical instabilities that result from the formation of multiple gas giants in a short range of distance from the star. Also, ice giants ($m\sim10M_\oplus$) do not survive in any of these simulations: they either became gas giants or were scattered out of the system by other giants. On the other hand, the results with $N_\text{ini}=2000$ and $5000$ do not show any qualitative difference from the presented results with $N_\text{ini}=1000$ (Fig. \ref{fig:2k_5k_mig_off_IGM16}).

\subsubsection{Pebble accretion efficiency $\epsilon_{\text{PA}}=\epsilon_\text{OL18}$} \label{sec:ol18}
Figure \ref{fig:1k_mig_off_OL18} shows the results for $N_\text{ini}=1000$, $S_a=0$ and $\epsilon_{\text{PA}}=\epsilon_\text{OL18}$. Compared to the results for $\epsilon_{\text{PA}}=\epsilon_\text{IGM16}$, the growth by pebble accretion is generally slower, but still rapid. Some planetesimals grow by up to about 2 orders of magnitude in mass in the first 0.1 Myr and massive cores ($m\sim M_\oplus$) are formed at 0.75 Myr. At 2.5 Myr, the massive cores in the inner part of the disc ($\sim5-10$ au) have reached the local pebble isolation mass and gas accretion begins with less than $\sim 10$ bodies having gained mass between the $\sim10M_\oplus$ cores and the initial planetesimals. In the previous simulations (Fig. \ref{fig:1k_mig_off_IGM16}), this stage is reached at 0.75 Myr. This delay is caused by the change in the adopted pebble accretion efficiency $\epsilon_{\text{PA}}$, where $\epsilon_\text{IGM16}$ is more efficient than $\epsilon_\text{OL18}$ as also shown in \cite{MatsumuraSoko2021}. A comparison between the two efficiency prescriptions and the consequences are further discussed in Sect. \ref{sec:dis_pa}. A more distinct dichotomy in mass is produced with $\epsilon_{\text{PA}}=\epsilon_\text{OL18}$ as shown by comparing the differential mass distribution in Fig. \ref{fig:1k_mig_off_IGM16} for 0.75 Myr and that in Fig. \ref{fig:1k_mig_off_OL18} for 2.50 Myr. A more significant number of planetesimals has reached $\sim10^{-3}M_\oplus$ in the former case while a sharper cut near the upper end ($\sim10^{-4}M_\oplus$) of the initial distribution is shown in the latter case. At this stage, the intermediate-mass bodies between these two groups, which have mass of about $10^{-5}-10^{-1}M_\oplus$, are generally dynamically colder, as shown by the $e$--$m$ and $i$--$m$ plots. As the simulations continue to 4.00 Myr, some bodies have become gas giants in the inner part of the disc, with some bodies of $\sim1-10M_\oplus$ residing outside of 10 au, in contrast to the results shown in Fig. \ref{fig:1k_mig_off_IGM16} at the same time.

At the end of the simulations, one to two gas giants and one to two ice giants are formed as well, which is the closest set of simulations in the work to reproduce the Solar System's giant planets. A significant number of the initial planetesimals remain, especially in the outer part of the disc at around 20 au. This is distinct from the results with $\epsilon_{\text{PA}}=\epsilon_\text{IGM16}$, where no ice giants are formed and most of the initial planetesimals have been scattered at the end of the simulations, probably due to the higher number of gas giants. Nonetheless, the locations of the leftover bodies still vary greatly across the simulations, so that the stochastic nature of the system remains.

Figure \ref{fig:2k_mig_off_OL18} shows the results for $N_\text{ini}=2000$ instead with the same pebble efficiency prescription. Compared to Fig. \ref{fig:1k_mig_off_OL18}, the differential mass distribution shows that the massive tail extends for about twice as high in $m$. This leads to the formation of more massive cores in the subsequent evolution of the simulations. At the end of the simulations, more gas giants and fewer ice giants are formed in this case. Only two out of the five simulations has one to four ice giants, while this class of bodies is absent in the rest of the simulations. With $N_\text{ini}=5000$, shown in Fig. \ref{fig:5k_mig_off_OL18}, only one simulation contains an ice giant at the end, which instead is located in the inner part of the disc at about 6 au. Here, we find a dependence on the value of $N_\text{ini}$, which is not present when $\epsilon_{\text{PA}}=\epsilon_\text{IGM16}$ (Sect. \ref{sec:igm16}). This is likely caused by the difference in the rate of pebble accretion, which is further discussed in Sect. \ref{sec:dis_pa}.

\subsection{Simulations with planet migration ($S_a=1$)}\label{sec:migon}
Figure \ref{fig:1k_mig_on_IGM16} shows the results for $N_\text{ini}=1000$, with migration $S_a=1$ and $\epsilon_{\text{PA}}=\epsilon_\text{IGM16}$. The snapshots of the $m$--$a$ distribution show that once the cores reach $\sim M_\oplus$, they migrate inwards rapidly, even though $\alpha_{\text{turb}}/\alpha_{\text{acc}} = 0.1$. For the massive cores that grow from planetesimals in the inner part of the disc, they have moved out of the simulation domain before runaway gas accretion occurs. For the massive cores that remain by the end of the simulations, the depletion of the gas disc stops both the migration as well as gas accretion. As a result, only cores of a few $M_\oplus$ are formed and survive in the simulations. A large fraction of the initial planetesimal population remains at the end as they are not scattered due to the absence of giant planets. Similarly, Fig. \ref{fig:1k_mig_on_OL18} shows the results for $\epsilon_{\text{PA}}=\epsilon_\text{OL18}$ with $S_a=1$ where only cores of a few $M_\oplus$ are formed and survive. These cores are slightly less massive in this case compared to Fig. \ref{fig:1k_mig_on_IGM16}. The results with $N_\text{ini}=2000$ and $5000$ do not show any qualitative difference from this results with migration in effect. Since the massive cores migrate rapidly and none reach the runaway gas accretion phase by the end of the simulation, the dependence on $N_\text{ini}$ shown in the case without planet migration for $\epsilon_{\text{PA}}=\epsilon_\text{OL18}$ (Sect. \ref{sec:ol18}) is no longer present in this case here.

\section{Discussion}\label{sec:dis}
\subsection{Pebble accretion efficiency}\label{sec:dis_pa}
Pebble accretion has been shown by the results of our model (Sect. \ref{sec:results}) to be a promising way to grow planetesimals efficiently such that massive cores of $\sim 10 M_\oplus$ can form well before disc dispersal and accrete gas to become giant planets. Nonetheless, forming giant planets analogous to those in the Solar System still requires further modifications to the model. In the presented results without planet migration ($S_a=0$), ice giants are formed only in the simulations with the pebble accretion efficiency prescription by \cite{LiuBeibei2018} and \cite{OrmelChrisW2018}, i.e. $\epsilon_{\text{PA}}=\epsilon_\text{OL18}$, as presented in Sect. \ref{sec:ol18}. The ice giants in these simulations stop accreting gas because by the time they are massive enough to accrete a gaseous envelope the gas disc is dispersed. In contrast with $\epsilon_{\text{PA}}=\epsilon_\text{IGM16}$, as shown in Sect. \ref{sec:igm16}, massive cores of $\sim10M_\oplus$ are formed much earlier and the giant planets have enough time to reach the prescribed final mass (Sect. \ref{sec:ga}) before disc dispersal. This shows that the timing of the formation of the massive cores and the start of gas accretion plays an important role in the final architecture of the planetary system.

As noted by \cite{MatsumuraSoko2021}, $\epsilon_\text{OL18}$ is generally a few times less efficient than $\epsilon_\text{IGM16}$ for the adopted value of $\alpha_\text{turb}$. And, in the present work, the simulations begin with a mass spectrum of planetesimals which spans over two decades in mass, up to $10^{-4}M_\oplus$, instead of lunar-mass embryos. This demonstrates the effect of the pebble accretion onset mass and the effect of viscous stirring on pebble accretion efficiency more clearly as discussed in the following.

\subsubsection{Pebble accretion onset mass}
First, we focus on the limit that the eccentricity $e$ of the pebble-accreting body is much lower than the midplane pressure gradient parameter $\eta\sim10^{-3}$. This is also an assumption held by \cite{IdaS.2016} in the derivation of the pebble accretion efficiency. Since we are considering the start of pebble accretion, the mass of the body is generally small and pebble accretion typically operates in the Bondi regime. In this case, the pebble relative velocity is determined by the headwind. For a high pebble relative velocity, the pebble encounter time is shortened so that pebbles may not be deflected enough from the gas streamline and not have enough time to settle onto the planetesimal. As such, the accretion is no longer in the settling regime. This reduction effect is captured in the pebble accretion efficiency prescription by \cite{IdaS.2016} as well as that by \cite{LiuBeibei2018} and \cite{OrmelChrisW2018} but in slightly different manners.
	
\cite{IdaS.2016} adopt the reduction factor for the cross section in the settling regime of pebble accretion proposed by \cite{Ormel2012}. This reduction factor is given by
\begin{equation}
    \kappa_\text{IGM16}=\exp\left[-\left( \frac{\text{St}}{\min(2,\text{St}_{\text{crit}})} \right) ^{0.65} \right] \label{eq:kappaida}
\end{equation}
with the critical Stokes number of pebble
\begin{equation}
    \text{St}_{\text{crit}}=\frac{4m}{\eta^3 M_*}.
\end{equation}
A similar reduction factor is also found in \cite{LiuBeibei2018}, which is given by
\begin{equation}
    \kappa_\text{OL18}=\exp\left[-\frac{1}{2}\left( \frac{\Delta v}{v_\text{crit}} \right) ^{2} \right] \label{eq:kappaOL}
\end{equation}
with the pebble relative velocity $\Delta v$ and the critical relative velocity
\begin{equation}
v_\text{crit} = \left( \frac{m}{M_*\text{St}}\right) ^{1/3}v_K.
\end{equation}
In the head wind regime, $\Delta v=\eta v_K$, and, with Eq. (\ref{eq:kappaOL}), the reduction factor can be expressed as
\begin{equation}
    \kappa_\text{OL18,hw}\approx\exp\left[-\left( \frac{\text{St}}{0.707\cdot\text{St}_{\text{crit}}} \right) ^{2/3} \right]
\end{equation}
for a more insightful comparison with $\kappa_\text{IGM16}$ in Eq. (\ref{eq:kappaida}). By inspection, the dependence on the planetesimal mass $m$ is virtually identical for both cases when $m\lesssim 2\times10^{-4}M_\oplus$ for $\eta=10^{-3}$, while a factor of about 0.707 is multiplied to $m$ for $\kappa_\text{OL18,hw}$. Figure \ref{fig:kappa} shows the values of $\kappa_\text{IGM16}$ and $\kappa_\text{OL18,hw}$ with an assumed St = 0.1 and $r=5$ au in our disc model. For $m\lesssim 10^{-5}M_\oplus$, $\kappa_\text{OL18,hw}$ is generally a few times smaller than $\kappa_\text{IGM16}$. This is in agreement with the findings by \cite{MatsumuraSoko2021} and the early stage of the presented simulation results.
 When the bodies are still dynamically cold, the growth by pebble accretion with $\epsilon_{\text{PA}}=\epsilon_\text{OL18}$ is generally slower.
 
While restricting the discussion in the headwind regime with small $e$, a pebble accretion onset mass $m_\text{PA,hw}$ can be defined \citep{VisserRicoG.2016,Ormel2017} by setting $\Delta v = v_\text{crit}$, which yields
\begin{equation}
    m_\text{PA,hw}=\text{St}\ \eta^3M_\ast. \label{eq:mpa}
\end{equation}
For $m=m_\text{PA,hw}$, this means $\kappa_\text{OL18}\approx0.61$ and $\kappa_\text{IGM16}\approx0.67$. Figure \ref{fig:mpa} shows a comparison of $m_\text{PA,hw}$ and the planetesimal gravitational mass of the adopted initial planetesimal mass function, $m_G$, at different locations of the disc. The increase with $r$ for $m_\text{PA,hw}$ is steeper than that for $m_G$. This is in agreement with the results that the growth by pebble accretion is faster in the inner part of the disc. Also, $m_G$ is about $5-10$ times smaller than $m_\text{PA,hw}$ from $5-20$ au. This means the massive tail of the planetesimal population overlaps with the mass range for the sharp cut off in the values of the reduction factors for both prescriptions ($\kappa_\text{IGM16}$ \& $\kappa_\text{OL18,hw}$) as shown in Fig. \ref{fig:kappa}. As a result, the randomness in the exact number of particles drawn near the top end of the distribution as well as that in their locations play a significant role to the final architecture of the modelled planetary systems.

This is more clearly shown by the difference in the results with $N_\text{ini}=\{1000,2000,5000\}$ while all have $S_a=0$ and $\epsilon_{\text{PA}}=\epsilon_\text{OL18}$. As the number of planetesimal increases, the largest drawn mass increases slightly as well due to the higher probability of getting at least one particle with such mass. This leads to an earlier formation of massive cores, which are more likely to become gas giants by the time of disc dispersal while fewer or no ice giants remain. Nonetheless, this effect is not observed with $\epsilon_{\text{PA}}=\epsilon_\text{IGM16}$ likely due to a generally more efficient pebble accretion such that gas accretion starts early for the massive cores with enough time to reach the mass of a gas giant even with $N_\text{ini}=1000$.
    
Although our results show an apparent dependence on the initial number of particles $N_\text{ini}$, we emphasize that this can be a result of a statistical artefact. With the adopted initial mass function by \cite{Abod2019}, as shown in Eq. (\ref{eq:dN}), there is no upper limit on the planetesimal mass. Although an artificial upper limit of 10 times of the characteristic mass is imposed, this limit has a negligible effect on the actual realized planetesimal populations, where only a number fraction of planetesimals of $\sim 8\times10^{-6}$ is lost. Therefore, the massive tail of the initial planetesimal population drawn in this manner has a dependence on the number of particles, which sets the normalization constant of the initial mass function. This means a physical upper limit of planetesimal mass \citep[e.g.][]{Gerbig2023} is needed to remove this artefact for future investigations. Nonetheless, our results show the upper end of the initial planetesimal population plays the most important role in growth by pebble accretion while the rest of the small planetesimals do not affect their growth significantly.

We note that in \cite{LambrechtsM.2012}, the transition mass of an embryo $m_\text{t}$ is defined as the mass at which the Hill radius is comparable to the Bondi radius, i.e.
\begin{equation}
    m_\text{t}=\sqrt{\frac{1}{3}}\frac{(\eta v_\text{K})^3}{G\Omega_K}\approx0.578\eta^3M_\ast.
\end{equation}
This mass is often adopted as the initial embryo mass in the works involving pebble accretion \citep[e.g.][]{Bitsch2015}. The value of $m_\text{t}$ is a few times larger than $m_\text{PA,hw}$ from Eq. (\ref{eq:mpa}) for $\text{St}=0.1$. This indicates these initial embryos can always grow by pebble accretion efficiently. The evolution from the point of planetesimal formation to the onset of pebble accretion is missing in this approach. We also note that the characteristic mass ($0.3M_G$) of the adopted initial planetesimal mass function is about an order of magnitude less massive than $m_\text{PA,hw}$, which is comparable to the value adopted by \cite{Coleman2021}. However, in the expression for $m_G$ in this work as shown by Eq. (\ref{eq:m_G}), the value of the small-scale diffusion parameter $\delta$ can be an order of magnitude larger or smaller than the adopted value \citep{Schreiber2018}. This translates to an even larger uncertainty in the initial planetesimal mass since $m_G\propto\delta^{3/2}$, which shall greatly change the results of our model and will require further investigations.

\subsubsection{Pebble accretion and dynamical heating}
However, as noted by \cite{IdaS.2016}, the assumption of small $e$ in the estimation of the pebble relative velocity only holds when $e<\eta\sim10^{-3}$. This condition breaks down quite early in the presented simulations, with a majority of the particles having $e$ exceeding $10^{-3}$ by 0.75 Myr in all the presented simulations. Multiple planet formation models \citep[e.g.][]{Levison2015,JangHyerin2022,LauTommyChiHo2022,Jiang2022} have shown the effect of increased pebble relative velocity due to dynamical heating on pebble accretion. Figure \ref{fig:kappa} also includes the general form of $\kappa_\text{OL18}$ with $\Delta v=\max(0.76e,\eta v_K)$ \citep{LiuBeibei2018} for $e=10^{-2}$, where the curve is shifted towards higher $m$ by more than two orders of magnitudes, i.e. a much larger $m$ is required for efficient pebble accretion. Therefore, it is likely an important feature of a realistic model to consider the effect of pebble accretion being interrupted when the eccentricities grow, especially in the context of planet formation where massive cores and giant planets are formed among planetesimals. However, once planet migration is in effect and cores of $\sim1-10M_\oplus$ are readily removed, they cannot continuously excite and eject the planetesimals. Pebble accretion in this case is not severely interrupted by dynamical heating as shown in the results (Sect. \ref{sec:migon}), so that both pebble accretion prescriptions yield more similar results at the end of the simulations when migration and removal is present.

We note that there are other works on the initial planetesimal mass function \citep[e.g.][]{Simon2016,Simon2017,SchaeferUrs2017,Gerbig2023}, and this topic remains an active field of research. Meanwhile, the outcome of the subsequent growth of the initial planetesimals is sensitive to their initial mass as well as the distribution. Also, we assume a planetesimal disc as a part of the initial conditions, but its formation is not investigated in this work, which is also an active field of research \citep[e.g.][]{Drazkowska2016,Carrera2017,Schoonenberg2018,Lenz2019,Lenz2020}. These parts of the model concerning the initial planetesimals require further investigations for a more robust planet formation model.

\subsection{Planet migration}
When planet migration is turned off in our model, i.e. $S_a=0$, the results with $N_\text{ini}=1000$ and $\epsilon_{\text{PA}}=\epsilon_\text{OL18}$ (Sect. \ref{sec:ol18}, Fig. \ref{fig:1k_mig_off_OL18}) show one to two gas giants and one to two ice giants beyond 6 au. This is in general agreement with \cite{Levison2012} in forming the giant planets in the Solar System without forming hundreds of massive cores in the process. In their work, planet migration is not considered either.

However, once planet migration is turned on in our model, i.e. $S_a=1$, the results (Sect. \ref{sec:migon}) show that cores of $\sim1-10M_\oplus$ rapidly migrate towards the inner part of the disc and many leave the simulation domain as a migration trap is not implemented at the inner edge of the disc. This is in agreement with previous works on planet formation that include planet migration \citep[e.g.][]{Cossou2014,Coleman2016a,MatsumuraSoko2017,JangHyerin2022}. Although the migration timescale in the high-mass regime in this work is already lengthened by setting a turbulent-$\alpha$ parameter $\alpha_\text{turb}$ that is only one-tenth of the classical $\alpha$ parameter $\alpha_\text{acc}$, it is still not enough to retain these massive cores at wide orbit in our model to form Solar-System-like giant planets. Further parameter search may be required to produce cold giant planets with planet migration in effect but the current results suggest that some massive cores are inevitably lost to the inner Solar System in the process as shown in other works \citep[e.g.][]{BitschBertram2019,MatsumuraSoko2021}.

Figure \ref{fig:tau_a} shows a heat map of the migration timescale $\tau_{a}$ in the $m$--$r$ space at $t=0.5$ Myr in our model. There is a region of rapid migration with $\tau_{a}\sim10^5$ yr for $m\sim1-10M_\oplus$ across the planetesimal disc. This is in agreement with the results that the massive cores have migrated significantly before runaway gas accretion can occur for them to enter the high-mass regime of migration where $\tau_{a}\sim10^6$ yr. For the surviving cores, migration only stops as the gas surface density becomes very low that slows down migration but this also terminates gas accretion as shown in the results. Also, it seems to be a general result that multiple massive cores ($\sim1-10M_\oplus$) inevitably enter the inner Solar System with a smooth disc model where migration trap is not present except at the inner edge of the protoplanetary disc. In contrast, other works \citep[e.g.][]{Coleman2016,LauTommyChiHo2022} have shown a possibility in retaining these cores at wide orbit due to the presence of a substructure in the gas disc. These findings and the recent observations of substructure in protoplanetary discs \citep[e.g.][]{Andrews2018,Long2018,Dullemond2018,Cieza2021} suggest that a substructure in the protoplanetary disc is a promising way to interrupt rapid migration and prevent the formation of super-Earths and hot Jupiters.

\section{Conclusions}\label{sec:concl}
This work attempts to form the giant planets of the Solar System in a smooth protoplanetary disc. An initial planetesimal disc is simulated with the parallelized $N$-body code SyMBAp with additional subroutines to include the effects of pebble accretion, gas accretion, and planet-disc interactions with the protoplanetary disc.

Our model starts from planetestesimals (each with $m\lesssim10^{-4}M_\oplus$) instead of planetary embryos ($m\sim10^{-2}M_\oplus$). In this work, we demonstrate the difference between the pebble accretion prescription by \cite{IdaS.2016} and that by \cite{LiuBeibei2018} and \cite{OrmelChrisW2018}. In \cite{IdaS.2016}, the pebble-accreting body is assumed to be in a circular orbit and the pebble relative velocity, which sets the pebble encounter time, is set by the headwind in the disc. In contrast, \cite{LiuBeibei2018} and \cite{OrmelChrisW2018} do not hold this assumption and consider the relative velocity due to eccentricity and inclination. In the case that the number of embryos is small and they are well above the pebble accretion onset mass both prescriptions give similar results, as noted in \cite{MatsumuraSoko2021}. However, in a planetesimal disc, viscous stirring becomes important and can effectively terminate growth by pebble accretion due to the increased pebble relative velocity and shortened pebble encounter time. This can occur when the inclinations of the bodies are small, and they are still well inside the pebble disc as also noted by \cite{LauTommyChiHo2022}. Therefore, to realistically model planet formation via pebble accretion starting from planetesimals, it is crucial to consider the reduced pebble encounter time due to dynamical heating.
    
When planet migration is not considered, our model can reproduce one to two gas giants and one to two ice giants beyond 6 au, which is analogous to the giant planets in the Solar System. However, we also note that the results have a dependence on the initial number of planetesimals. Further studies on the processes involved in planetesimals formation is required to construct a more realistic model.

Once planet migration is in effect, massive cores of about 10 $M_\oplus$ are readily removed as they migrate towards the inner boundary of the simulations. This shows that the formation of the giant planets in the Solar System requires an alternative and effective way to stop the migration of the first massive body formed before reaching the inner Solar System. Multiple works \citep[e.g.][]{Coleman2016,LauTommyChiHo2022} have demonstrated that pressure bump in the disc can act as a migration trap while some other works \citep[e.g.][]{Jiang2022,Chrenko2023} do not support this scenario. Further investigations are required to characterize the disc conditions that can retain massive planetary cores and allow the formation of cold gas giants.

\begin{acknowledgements}
    This work was supported in part by a postgraduate studentship (T.C.H.L.) and a seed fund for basic research (T.C.H.L and M.H.L.) at the University of Hong Kong and Hong Kong RGC grant 17306720 (T.C.H.L and M.H.L.). Parts of the computations were performed using research computing facilities offered by HKU Information Technology Services. T.C.H.L also acknowledges funding from the Deutsche Forschungsgemeinschaft (DFG, German Research Foundation) under grant 325594231. This research was supported by the Munich Institute for Astro-, Particle and BioPhysics (MIAPbP) which is funded by the Deutsche Forschungsgemeinschaft (DFG, German Research Foundation) under Germany´s Excellence Strategy – EXC-2094 – 390783311.
\end{acknowledgements}

\bibliographystyle{aa}
\bibliography{ms}

\begin{appendix}
\section{Additional figures}
\begin{figure}[h]
	\centering
	\resizebox{\textwidth}{!}{\includegraphics{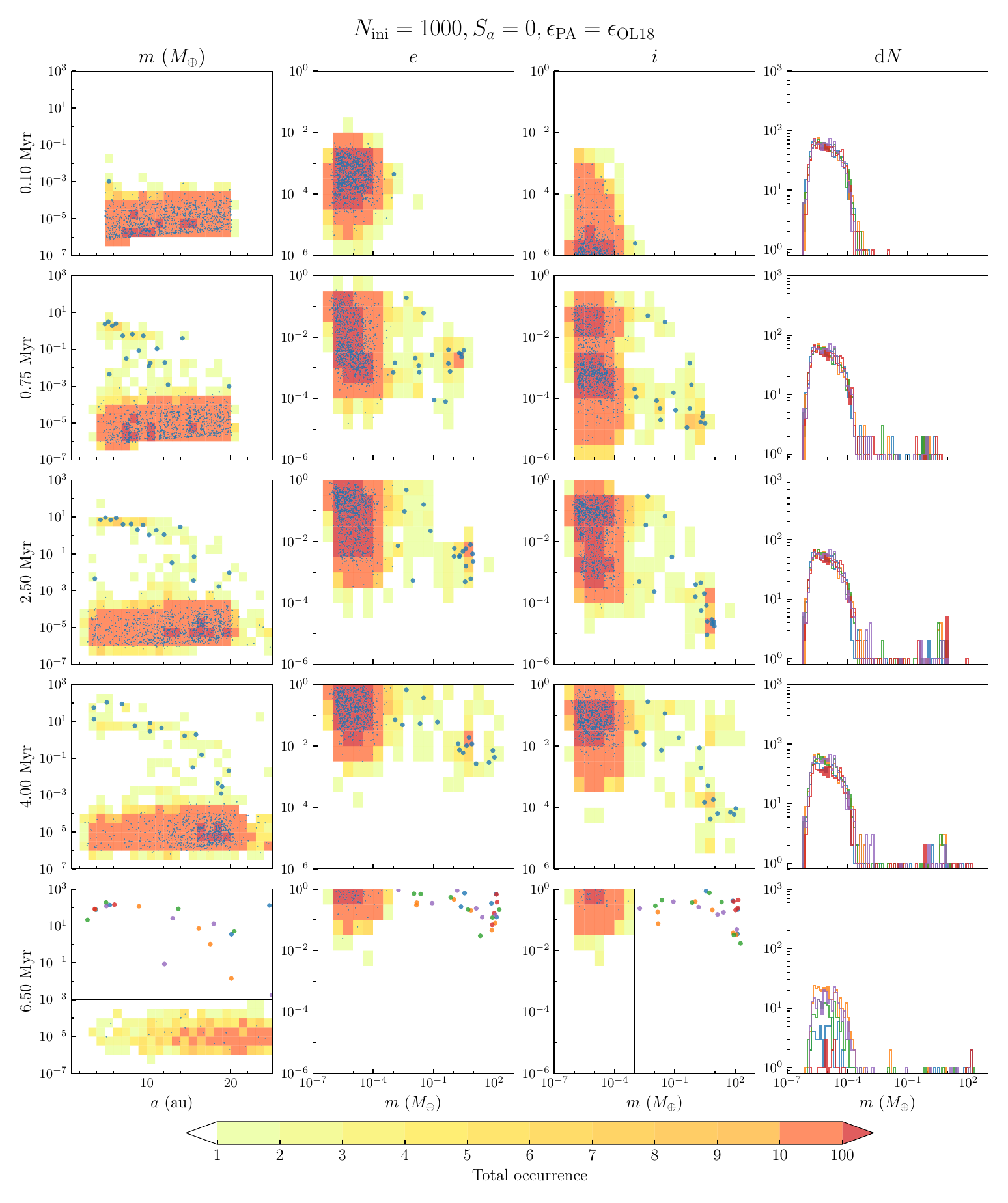}}
	\caption{Same as Fig. \ref{fig:1k_mig_off_IGM16} except $\epsilon_{\text{PA}}=\epsilon_\text{OL18}$.}
	\label{fig:1k_mig_off_OL18}
\end{figure}

\begin{figure*}
	\centering
	\includegraphics[width=\textwidth]{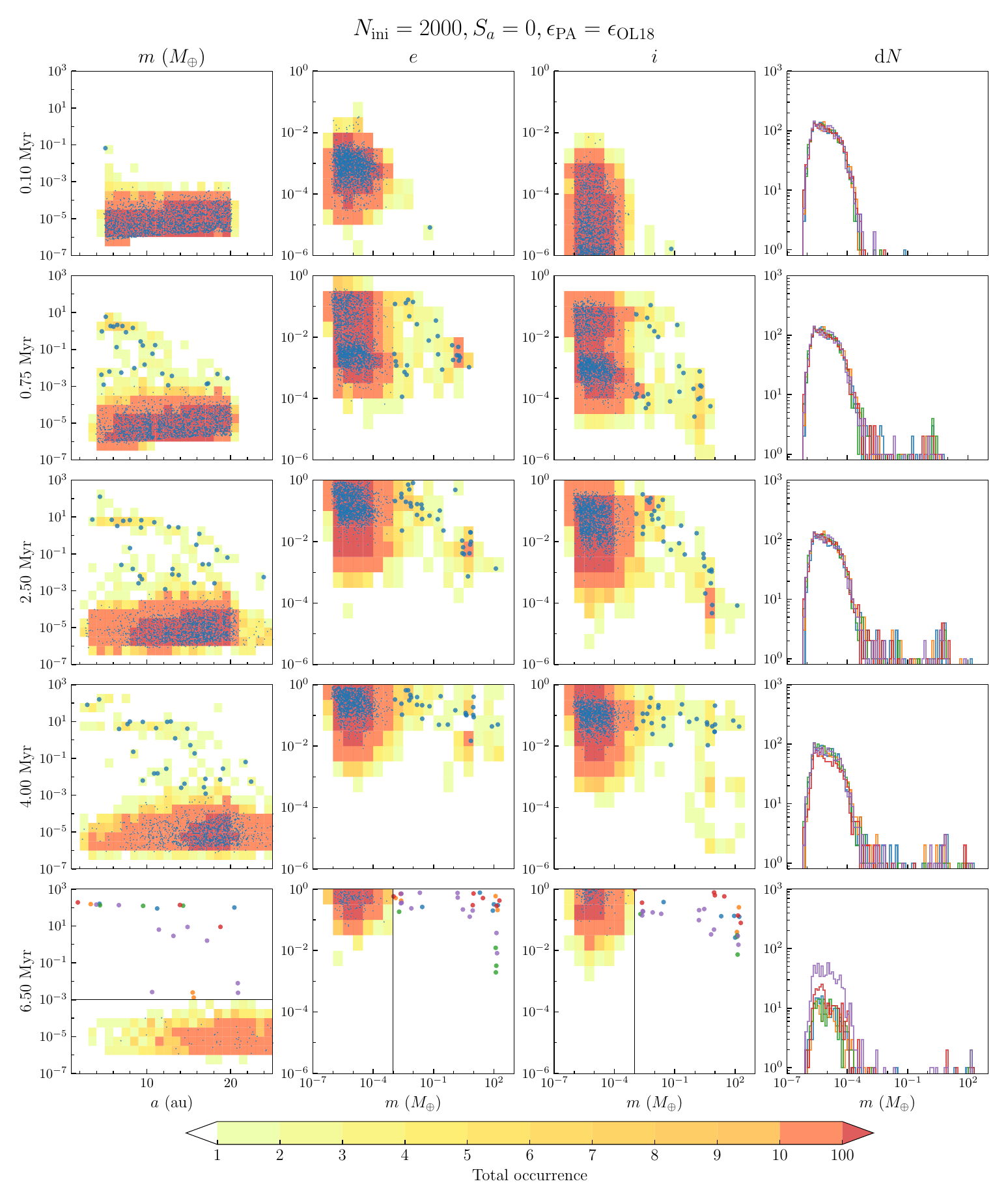}
	\caption{Same as Fig. \ref{fig:1k_mig_off_OL18} except $N_\text{ini}=2000$.}
	\label{fig:2k_mig_off_OL18}
\end{figure*}

\begin{figure*}
	\centering
	\includegraphics[width=\textwidth]{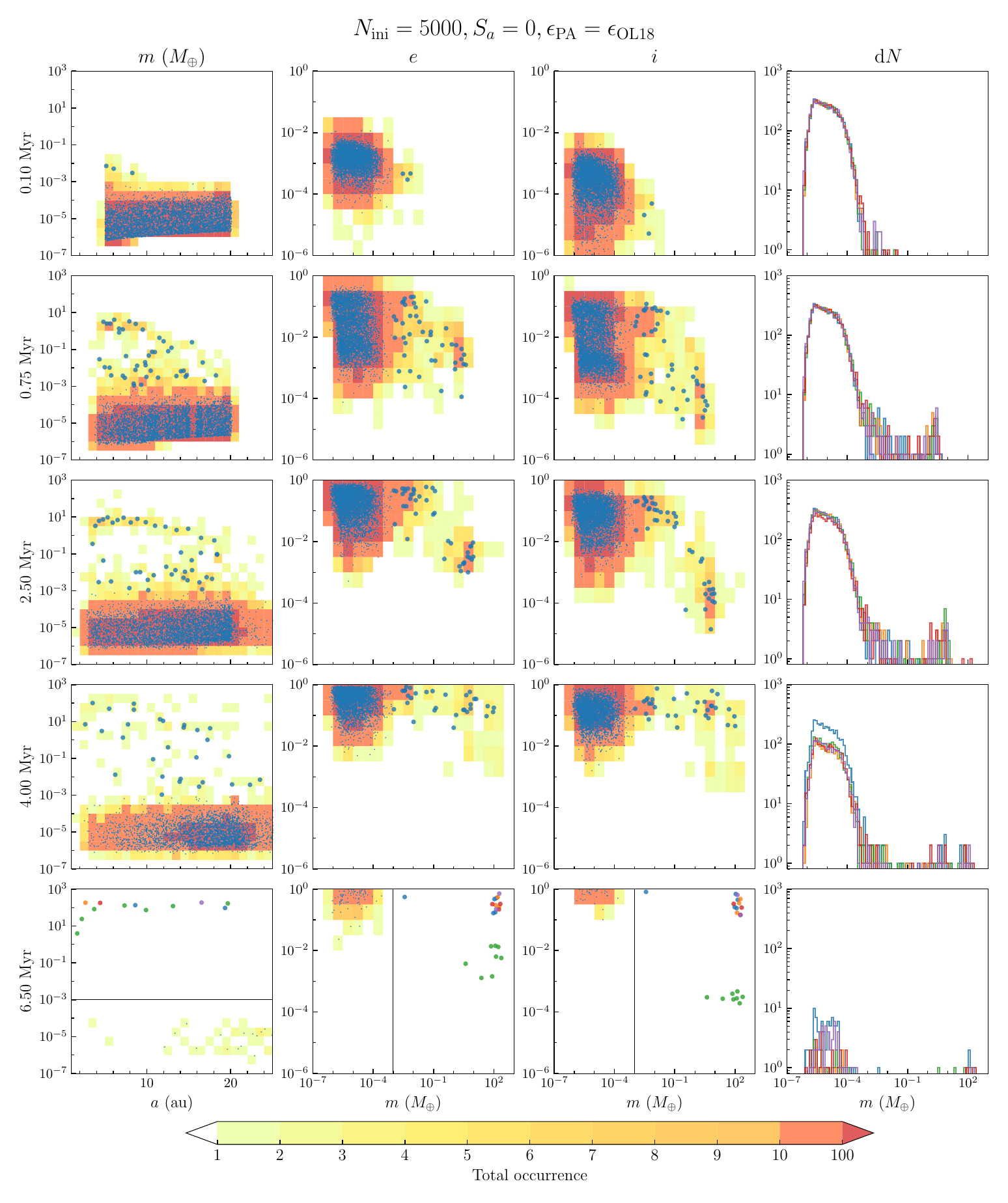}
	\caption{Same as Fig. \ref{fig:1k_mig_off_OL18} and \ref{fig:2k_mig_off_OL18} except $N_\text{ini}=5000$.}
	\label{fig:5k_mig_off_OL18}
\end{figure*}

\begin{figure*}
	\centering
	\includegraphics[width=\textwidth]{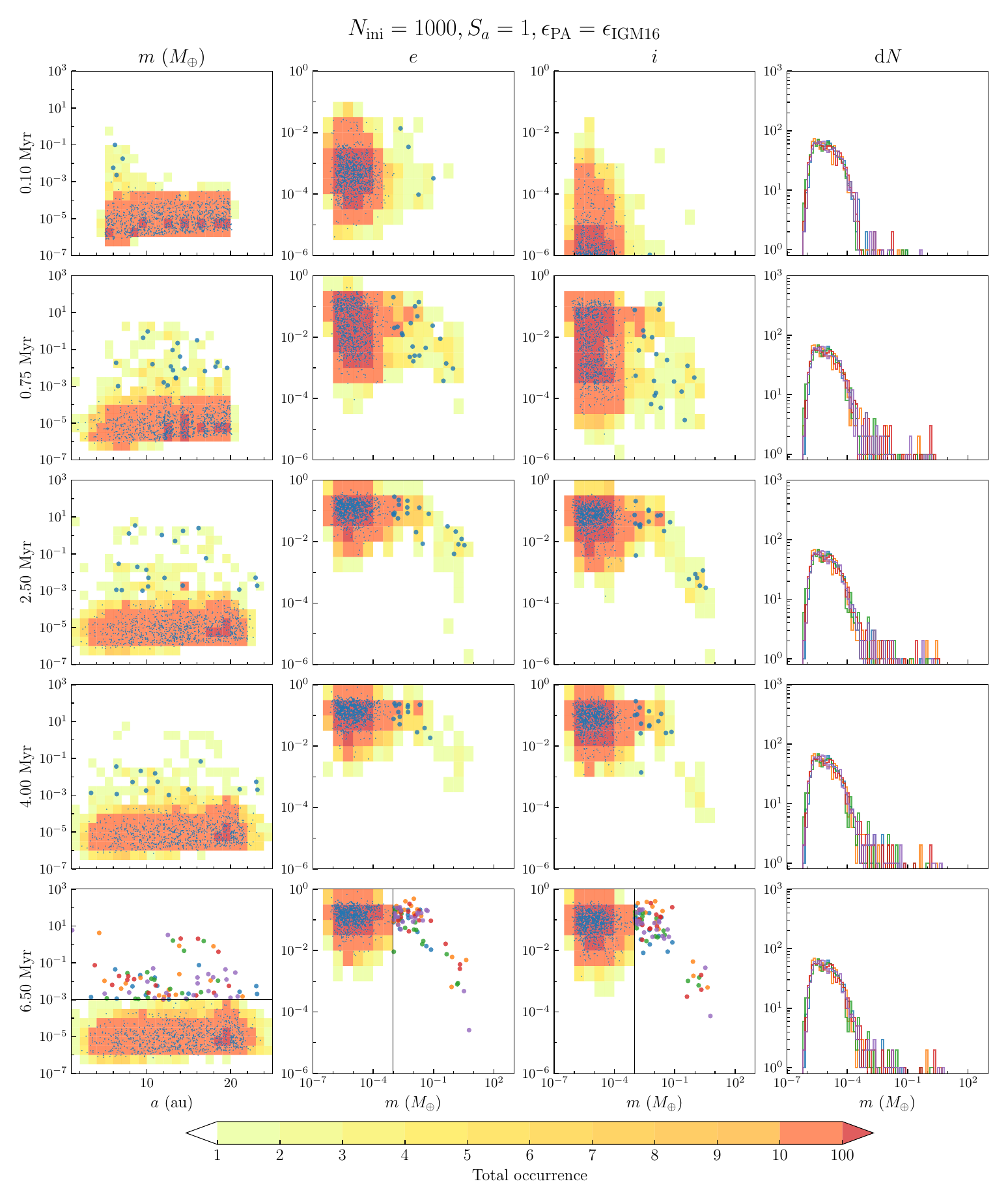}
	\caption{Same as Fig. \ref{fig:1k_mig_off_IGM16} except with migration turned on ($S_a=1$).}
	\label{fig:1k_mig_on_IGM16}
\end{figure*}

\begin{figure*}
	\centering
	\includegraphics[width=\textwidth]{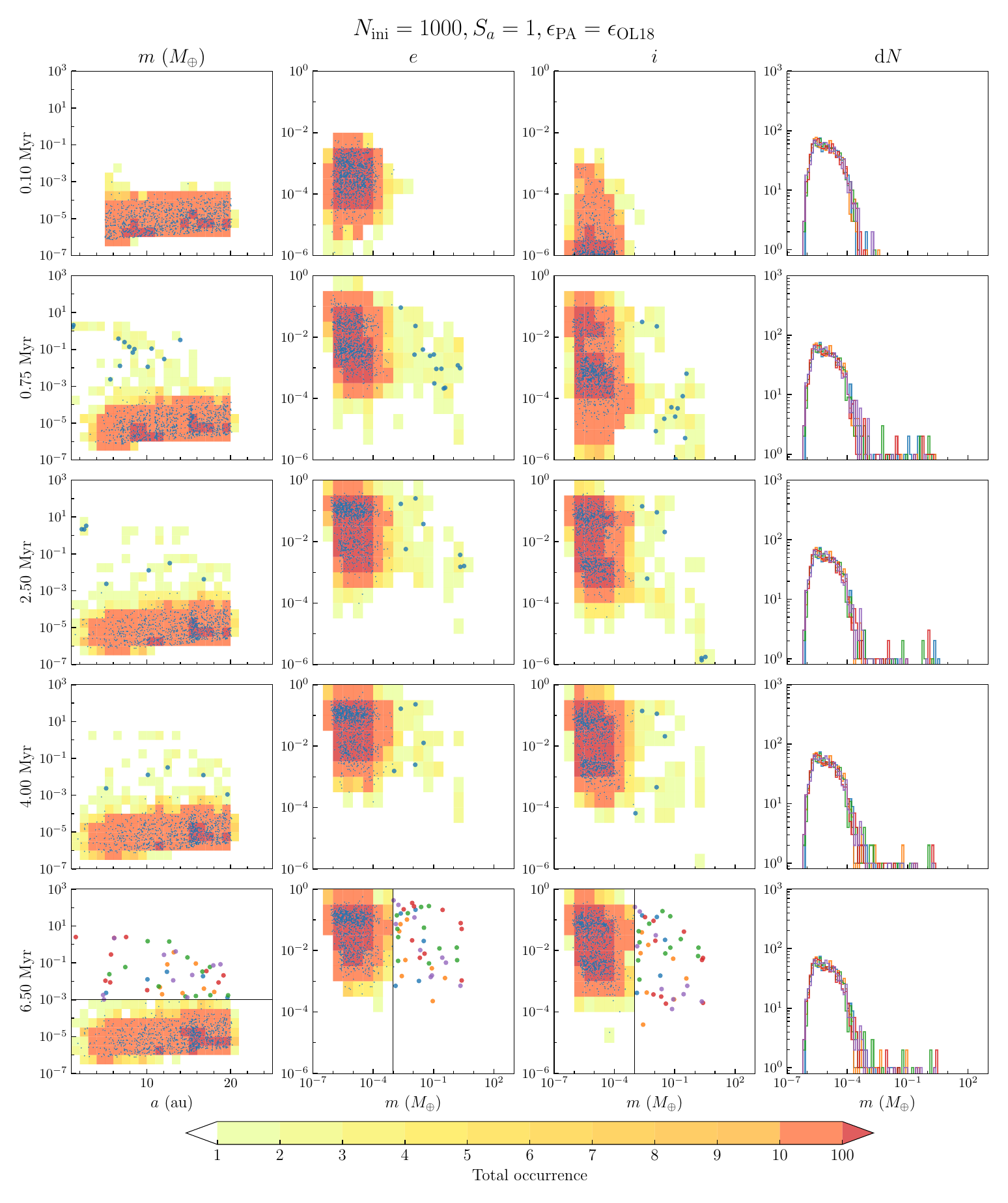}
	\caption{Same as Fig. \ref{fig:1k_mig_off_OL18} except $S_a=1$.}
	\label{fig:1k_mig_on_OL18}
\end{figure*}

\begin{figure}
    \centering
    {\graphicspath{{./fig/}} \input{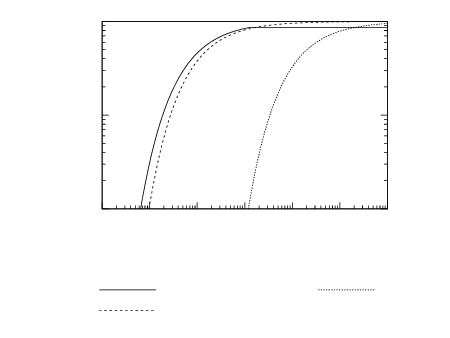}}
    \caption{Reduction factor $\kappa$ on the pebble accretion cross section in our disc model according to the prescription by \cite{IdaS.2016}, $\kappa_\text{IGM16}$ (solid line), a similar reduction factor by \cite{LiuBeibei2018} in the head wind regime with small $e$ and $i$ $\kappa_\text{OL18,hw}$ (dashed line), and that with $e=10^{-3}$ and small $i$ $\kappa_\text{OL18},e=10^{-2}$ (dotted line) as different mass $m$. The values of St = 0.1 and $r=5$ au are set for this estimation.}
    \label{fig:kappa}
\end{figure}

\begin{figure}
    \centering
    {\graphicspath{{./fig/}} \input{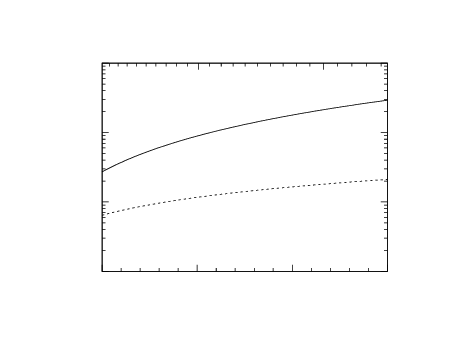}}
    \caption{Pebble accretion onset mass $m_\text{PA,hw}$ in the headwind regime (solid line) and the planetesimal gravitational mass $m_G$ (dashed line) at different locations of the disc $r$ with the corresponding value of $\eta$ shown in the upper $x$-axis. The values of $m_G$ are around the mass range for the steep cutoff in $\kappa$ shown in Fig. \ref{fig:kappa}.}
    \label{fig:mpa}
\end{figure}

\begin{figure}
\centering
{\graphicspath{{./fig/}} \input{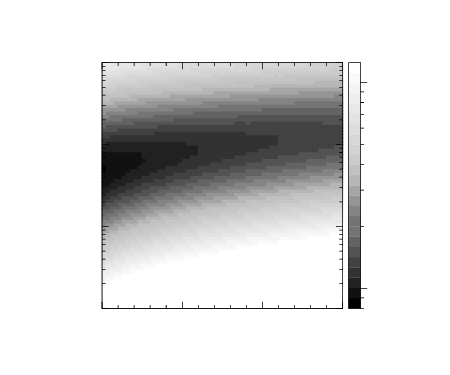}}
\caption{Heat map of the migration timescale $\tau_{a}$ in the $m$--$r$ space at $t=0.5$ Myr in our model. A region of rapid migration ($\tau_{a}\sim10^5$ yr) is presence for $m\sim1-10M_\oplus$ across the planetesimal disc ($r=5-20$ au).}
\label{fig:tau_a}
\end{figure}

\end{appendix}
	
\end{document}